\documentclass[journal=nanole,layout=twocolumn,articletitle=true]{achemso}
\setkeys{acs}{articletitle = true}

\usepackage[utf8]{inputenc}
\usepackage[T1]{fontenc}
\usepackage{amssymb}
\usepackage{color}

\author{Tobias Preis}
\affiliation{Institute of Experimental and Applied Physics, University of Regensburg, 93040 Regensburg, Germany}
\author{Sasha Vrbica}
\affiliation{Huygens-Kamerlingh Onnes Laboratory, Leiden University, 2333 CA Leiden, The Netherlands}
\author{Jonathan Eroms}
\affiliation{Institute of Experimental and Applied Physics, University of Regensburg, 93040 Regensburg, Germany}
\author{Jascha Repp}
\affiliation{Institute of Experimental and Applied Physics, University of Regensburg, 93040 Regensburg, Germany}
\author{Jan M. van Ruitenbeek}
\affiliation{Huygens-Kamerlingh Onnes Laboratory, Leiden University, 2333 CA Leiden, The Netherlands}
\email{ruitenbeek@physics.leidenuniv.nl}

\title[Diffusion of Co on graphene nanoribbons]{Current-induced one-dimensional diffusion of Co ad-atoms on graphene nanoribbons}

\begin{document}

\begin{abstract}
One-dimensional diffusion of Co ad-atoms
on graphene nanoribbons has been induced and investigated by means of scanning tunnelling microscopy (STM). 
To this end, the nanoribbons and the Co ad-atoms have been imaged before and after 
injecting current pulses into the nanoribbons, with the STM tip in direct contact with the ribbon.
We observe current-induced motion of the Co atoms along the nanoribbons, 
which is approximately described by a distribution expected for 
a thermally activated one-dimensional random walk. 
This indicates that the 
nanoribbons reach temperatures far beyond 100\,K, 
which is well 
above the temperature of the underlying Au substrate. 
This model system can be 
developed further for the study of electromigration at the single-atom level.
\end{abstract}

\vskip 2cm

Graphene nanoribbons (GNR) 
can be engineered with atomic perfection by on-surface synthesis 
from molecular precursors,\cite{Cai2010} 
offering a rich variety of fascinating electronic properties \cite{Rizzo2018,Groning2018,Slota2018,Nguyen2017,Jacobse2017}.
Their band gap can be tuned by the physical width of the ribbon,
such that they represent nearly ideal model systems for investigating one-dimensional (1D) electron transport \cite{Koch2012,Jacobse2018}. 
GNR can be intrinsically doped by adsorption onto a surface with a different work function
\cite{Li2013, Ijaes2013, vanderLit2013, Baringhaus2013}, or the doping can be controlled by substituting 
carbon atoms in the lattice \cite{Kawai2015, Carbonell-Sanroma2017b}. 
Alternatively, dopants can be deposited onto the GNR,\cite{Chen2008,Elias2020} but little is known of how a current flow through GNR may affect adsorbed dopants. 

Here, we demonstrate that high current densities will drive dopant atoms to diffuse, at bath temperatures for which thermal diffusion is inhibited. 
Interestingly, the diffusion is confined to the GNR, rendering GNR unique model systems for studying atomic diffusion in one dimension. 
Specifically, Co ad-atoms on top of GNR, which themselves are adsorbed on Au(111),
are investigated by means of low-temperature scanning tunneling microscopy (STM).
We inject large current densities into the GNR, with the STM tip in direct contact with the GNR,
at various lateral distances from the targeted Co ad-atoms. 
Thus, driving diffusion of the ad-atoms,
we find that nearly all Co ad-atoms remain confined to the GNR, and diffuse along the ribbon. 
We analyze the statistical distribution of current-induced lateral displacements, 
exhibiting a non-directional hopping that is similar to what would be expected for a thermally-driven process. 
We anticipate that the system can be further developed as a model system for study the electromigration at the atomic scale, and testing fascinating predictions.\cite{Dundas2009,Lu2010}   

The experiments were carried out using a low-temperature 
ultra-high vacuum STM (Createc design with modifications). 
Unless specified differently, the experiments were performed 
at the base temperature of 7--10 K
and STM images were acquired with 
a set-point tunneling current of $I = 2$\,pA at a bias voltage $V = \pm 100$\,mV. 
Bias voltages refer to the sample voltage with respect to the tip.

\begin{figure}[t!]
\includegraphics[width=\linewidth]{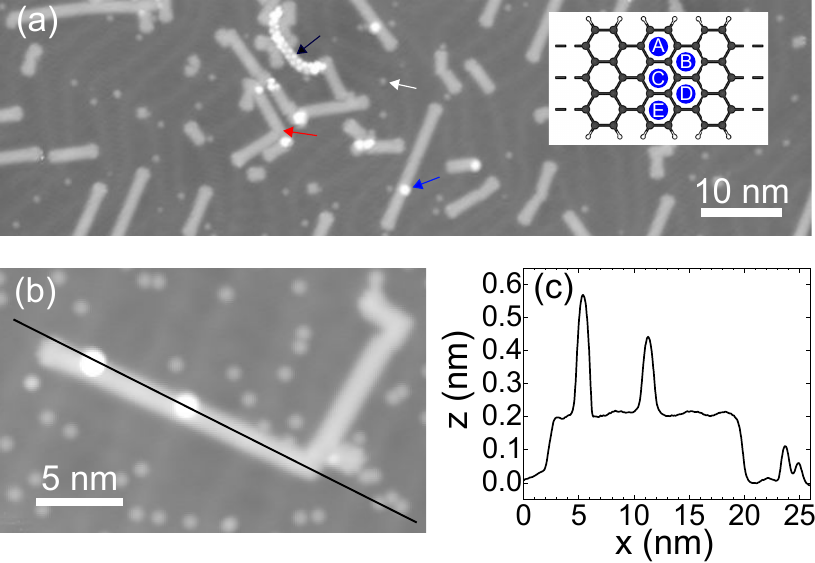}
\caption[Structure of GNR]{
				Overview of GNR and co-adsorbed Co ad-atoms on Au(111). 
		               (a) Large-area STM image of the sample after cobalt ad-atom deposition. 
		               Some GNR are not fully flattened out from dehydrogenation (black arrow)
		               and a few GNR are kinked (red arrow) or deviate otherwise from their ideal 1D structure.
		               Most Co ad-atoms are located on clean areas of Au (white arrow), while a few 
		               are located on top of GNR (blue arrow).
		               The inset shows the geometric structure of the GNR along with the five 
		               possible ring-centered adsorption sites per unit cell.		               
		               (b) STM image of a GNR decorated by two Co ad-atoms.  
		               (c) Line profile along one arm of the kinked GNR, indicated by the black line in (b) 
}		\label{Fig:structure_GNR}
\end{figure}

A Au(111) single-crystal surface was cleaned by Ne$^{+}$ ion sputtering, followed by thermal annealing at 550\,$^{\circ}$C. 
GNR on Au(111) were synthesized from 10,10'-dibromo-9,9'-bianthryl precursors 
on the surface, following the method introduced by Cai {\it et al.} in Ref.~\citenum{Cai2010}. 
The resulting structure is 7-aGNR, that is, 7-carbon-dimers-wide ribbons 
with armchair edges. 
After cooling down the sample and transferring it into the STM, Co atoms 
were co-adsorbed onto the surface by electron beam evaporation. 
In STM images, Fig~\ref{Fig:structure_GNR}(a), we find a significant fraction
of the Co atoms adsorbed on top of the GNR. 

As shown previously,\cite{Cai2010, Dienel2015} GNR fabricated by this method 
have a uniform width, but can be of various length, they can have kinks, and different GNR can touch or overlap.
Whereas Co on Au(111) has an apparent height of 0.1 nm in the STM images, Co ad-atoms on a GNR show typically an apparent height of 0.23 nm.
We tentatively attribute this difference to a reduced electronic coupling of the Co ad-atoms when being adsorbed on a GNR as compared to direct adsorption
on Au and note that metal ad-atoms adsorbed on ultrathin insulating layers also show large apparent heights in STM.\cite{Repp2004b}  
Occasionally, some Co-atom-related protrusions on GNR appear even larger -- about 0.35 nm high. 
The larger protrusions may be due to Co dimers,
or to Co monomers adsorbed at a different site,  
leading to a different electronic state of the atom\cite{Virgus2014}. 
In our study, we considered Co monomers only and disregarded the larger protrusions.

At the base temperature of $\sim 7$\,K the Co atoms and the GNR remain stationary. 
It is even possible to pick up the edge of a GNR by the STM tip, and drag it laterally over the surface, 
while the Co ad-atoms remain at fixed positions on this GNR, providing strong evidence that the Co ad-atoms are indeed adsorbed on top of the GNR. 
Although the atomic structure of the GNR lattice is not resolved in STM at the scanning parameters applied here, 
the preferred adsorption sites can nonetheless be deduced.
From suitable fitting procedures, the center position of Co ad-atoms as well as the central axis of a GNR can be extracted.
From the analysis of the relative position of the center positions  
we find 
only adsorption at positions A, B, D and E (see inset Fig~\ref{Fig:structure_GNR}(a)). 
This observation is consistent with Co preferentially adsorbing at the centers of the outer carbon rings of the GNR,
in agreement with Ref.~\citenum{Sevincli2008}.

\begin{figure}[t!]
	\begin{center}
		\includegraphics[width=\linewidth]{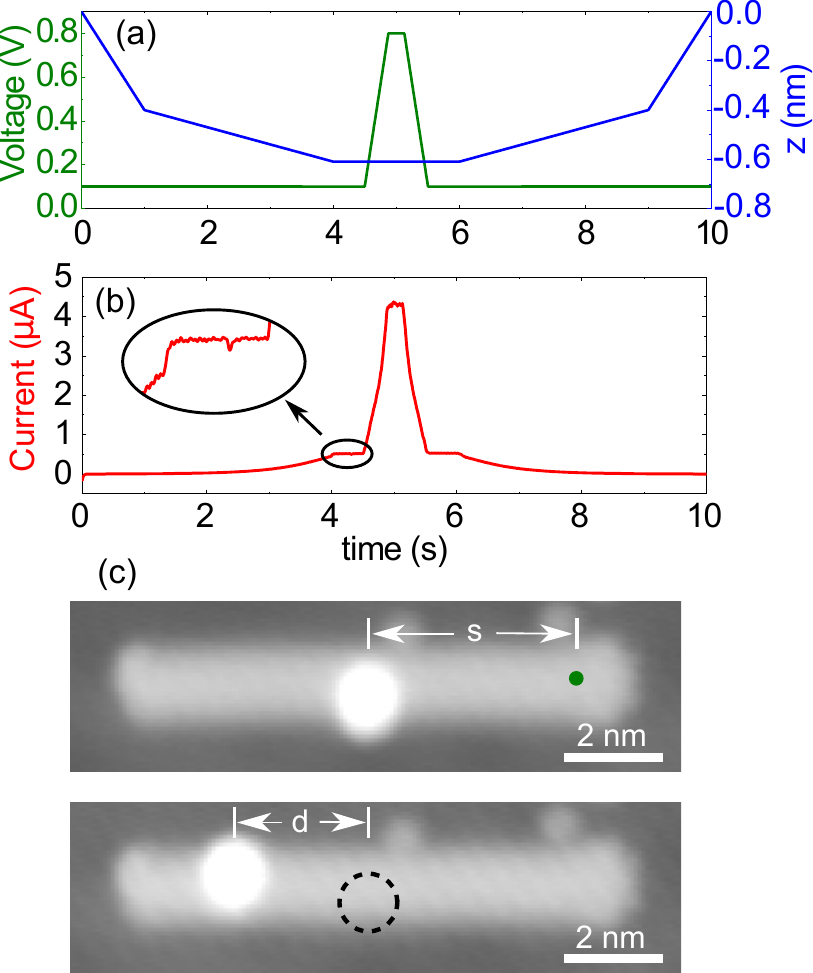}
		\caption[Pulsing procedure]{
		Procedure of current-pulse injection.
			(a)  The blue curve indicates the tip approach towards the GNR. A voltage pulse is applied in contact, here 0.8\,V for a duration of 0.5\,s (green curve). 
			(b) Tunneling current recorded during the pulse. An abrupt increase indicates the jump to contact (see zoom-in).
			(c) STM images of a GNR before (top) and after (bottom) the voltage pulse for $V_{\rm max}$ = 0.8\,V. 
			The green dot on the GNR marks the tip position for the applied voltage pulse, at a distance $s$ from the ad-atom's initial position. 
			After the pulse, the Co ad-atom has moved away from the tip over a distance $d$, as observed in the lower image. }
		\label{pulsing_scheme}
	\end{center}
\end{figure}

We now turn to the current injection with the STM tip into GNRs to induce motion of the ad-atoms.
To this end, the STM tip is positioned above a suitable GNR at a lateral distance $s$ from a Co atom, as shown by the green dot in Fig.~\ref{pulsing_scheme}(c), top panel. 
Next, the feedback is switched off and the tip is lowered ($z$ decreases) as shown by the blue curve in Fig.~\ref{pulsing_scheme}(a).
The tunneling current, Fig.~\ref{pulsing_scheme}(b), rises exponentially with decreasing tip-sample distance 
until a kink is observed in most current traces, indicating a jump to point contact \cite{Untiedt2007,Kroeger2007}.
The tip approach is continued for another 0.02--0.03\,nm  in order to ensure stable contact. 
When in contact, the sample voltage is linearly ramped up, briefly maintained at $V_{\rm max}$
and then linearly ramped down to the initial imaging conditions, over a total duration of 1\,s,
see Fig.~\ref{pulsing_scheme}(a). 
After this procedure, the tip is retracted, the feedback switched back on, and another 
STM image is taken and compared to the previous one; an example is shown in Fig.~\ref{pulsing_scheme}(c).
By comparing the images before and after the pulse the displacement $d$ of the Co atom can be determined.
A displacement away from the tip is registered with a positive sign whereas a displacement towards the tip is counted with a negative sign.
{ In the following, we occasionally refer to current pulses injected into the ribbons, but we emphasize here that the voltage is the controlled parameter. }

We have applied over 900 pulses on GNRs, and recorded a total of 1430 events (noting that a single pulse may affect multiple Co ad-atoms on a GNR).
Remarkably, we find that after application of a voltage pulse roughly half of the Co atoms investigated are displaced to a new position on the same GNR, see Table~\ref{Tab:pulsing_statistics}. 
Atoms can be displaced both ways, namely towards or away from the tip, for either voltage polarity, 
and the average displacement is quite significant, averaging at 3--4 graphene hollow sites. 
Only in very few cases Co atoms were found to be moved from the GNR to the bare Au surface. 
Finally, roughly 7\% of the atoms disappeared from the image after the pulse. 
Since these atoms were not found in large area surface scans we assume that they must have been picked up by the tip. 

These experimental observations are analyzed in more detail below.
We will restrict the analysis to experiments for voltage pulses $|V| \geq 600$\,mV; for smaller voltage values we do not have sufficient statistics. 
We also exclude data for which the tip-sample resistance remained larger than 330 k$\Omega$ during the contacting procedure.
Atoms located at the end points of the GNR or at kink sites, where the potential landscape may significantly differ, are disregarded. 
Larger protrusions, most probably arising from clusters of Co atoms were also excluded. 
\begin{table*}
	\centering
	\begin{tabular}{|l|l|l|}
		\hline
		{\bf Response to voltage pulses of Co ad-atoms on GNRs}	& {\bf number} &  {\bf \% of total } \\  \hline \hline
		Total number of Co atoms traced	& 1430 & 100\%\\  \hline 
		Number of Co atoms that did not move 	& 736 & 51.5\% \\ \hline
		Number of displaced Co atoms (including events below)	& 694 & 48.5\%  \\ \hline
		\hspace{0.5cm} Co moved along GNR	& 536 & 37.5\% \\ \hline
		\hspace{0.5cm} Co dropped to Au surface	& 6 & 0.4\% \\ \hline
		\hspace{0.5cm} Co moved to other edge (between sites A and E)	& 54 & 3.8\% \\ \hline
		\hspace{0.5cm} Co atom disappeared (pick up by STM tip)	& 98 & 6.8\%  \\ \hline
	\end{tabular}
	\caption[Statistics for Co motion after a voltage pulse.]{{Statistics for Co motion after a voltage pulse.} }
	\label{Tab:pulsing_statistics}
\end{table*}

\begin{figure}[b!]
	\begin{center}
		\includegraphics[width=\linewidth]{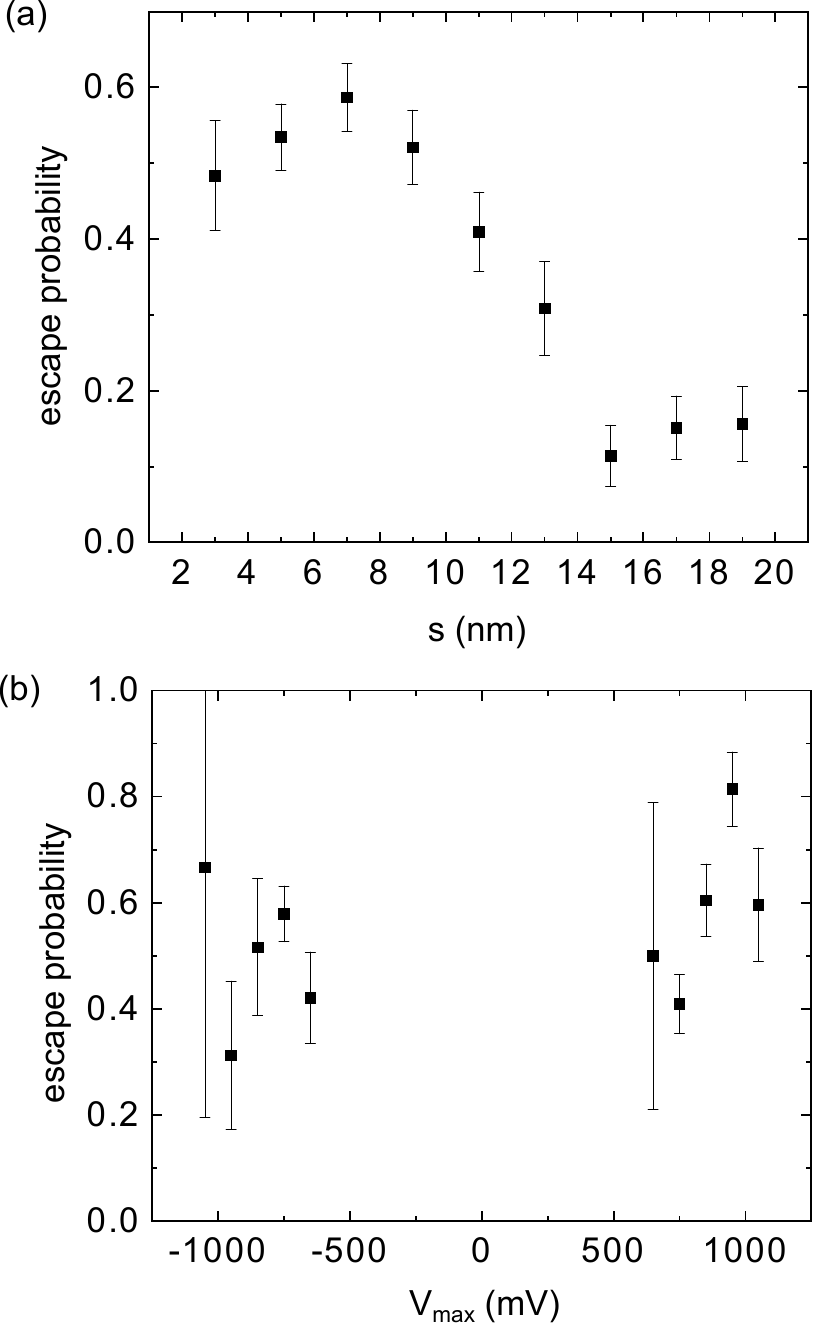}
		\caption[Escape probability]{
		Escape probability (a) as a function of initial tip-ad-atom separation, and (b) as function of pulsing voltage, for pulses with $s$ < 10\,nm. 
		Error bars show the standard deviation.}
		\label{Fig:escape_probability}
	\end{center}
\end{figure}

In Fig.~\ref{Fig:escape_probability}(a) we show the escape probability of Co ad-atoms on GNRs after a voltage
pulse, as a function of the initial tip -- ad-atom distance $s$. We define the escape probability as the ratio of the atoms 
that are displaced from their initial position after a pulse (excluding those that have presumably been picked up by the tip), to the total number of Co atoms traced. 
The data are grouped in intervals of initial distances $s$, 
and a data point at $s$ combines data for the interval from $s-1$ nm to $s+1$ nm. 

The escape probability is first roughly constant slightly above 50\%
up to a distance of $s = 10$\,nm, after which it drops gradually to a value of $0.1-0.2$, and becomes constant again
up to $s = 20$\,nm.
Clearly, atoms closer to the tip are more likely to be displaced. 
We did not investigate possible effects of the length of the GNR on the escape probability, nor effects due to the presence of kinks or overlaps. 
The initial distance $s$ does not influence the direction of motion. 

Figure~\ref{Fig:escape_probability}(b) shows the escape probability as a function of $V_{\rm max}$. 
For this analysis we used all data for $s < 10$\,nm, since the dependence on $s$ in this range is small.
Each data point represents an interval of 0.1\,V. 
Despite some scatter we can recognize a trend of increasing escape probability for increasing bias voltage.

\begin{figure}[t!]
	\begin{center}
		\includegraphics[width= \linewidth]{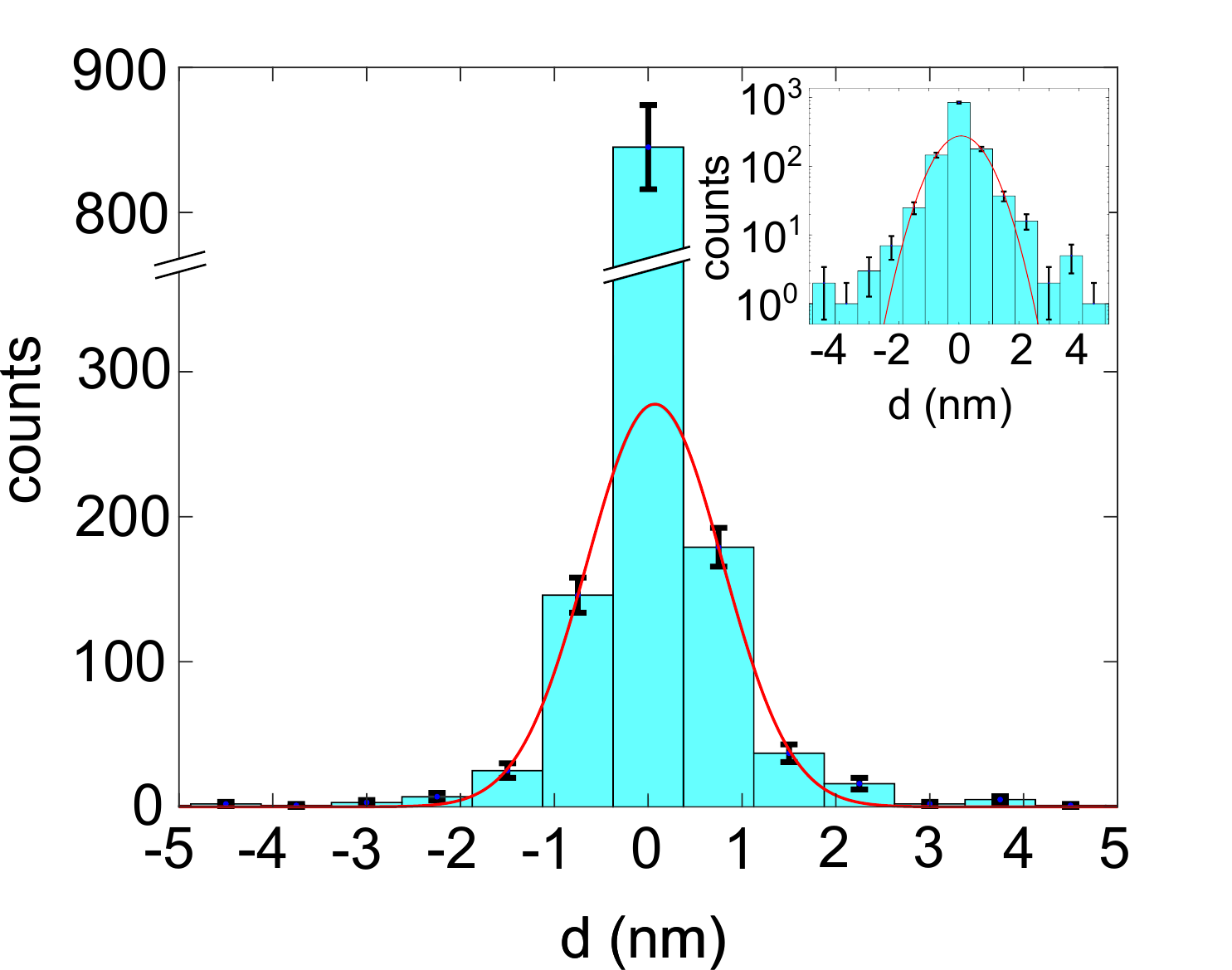}
		\caption[Escape probability]{Histogram of the distance $d$ traveled by Co ad-atoms along the GNR after a voltage pulse, with $t_p \simeq 0.5$\,s. 
		Positive (negative) $d$ indicates motion away from (towards) the tip, while $d=0$\,nm indicates Co 
		ad-atoms that did not move. The red curve shows a fit of the data to a 1D random walk
		probability distribution, { Eq.~(\protect\ref{eq.random-walk}) with $N \approx 13$,}  excluding the data points for $d=0$. 
		The inset shows the data on a semi-log scale, emphasizing the deviations to the fit for large displacements $d$. }
		\label{Fig:histogram}
	\end{center}
\end{figure}

Figure~\ref{Fig:histogram} shows a histogram of the distance $d$ by which ad-atoms are displaced as a result of  a pulse. 
The graph combines all initial distance values $s$ and all voltage pulse values $|V| \geq 0.6$\,V.  
The largest displacements observed are $\sim 8$\,nm in both directions. 
The red curve is a fit of the histogram, excluding the anomalous bin at $d=0$,  to 
$P(d)=A e^{{-(d-d_0)^2/w^2}}$. 
The center of the distribution lies at zero, to within the statistical accuracy, $d_0 = 0.1 \pm 0.8$\,nm, as expected
for a random walk probability distribution.
Analyzing the data separately for positive and negative voltage pulses we find $d_0^+ = 0.2 \pm 1.0$\,nm for $V>0$, and $d_0^- = 0.05 \pm 1.5$\,nm for $V<0$, showing that 
we do not resolve a preferential diffusion direction.  
From the width $w^2=1.0$\,nm$^2$ we find that the average number of hopping events during a pulse is about 13.

Although the fit appears to match the data reasonably well on a linear scale, the inset showing the data on a semi-log scale emphasizes deviations at larger distances.
The events with displacements by more than 3\,nm appear orders of magnitude more frequently than expected from the simple 1D random walk. We return to a discussion of these long tails, and the anomalously large peak at $d=0$, below.

During the experiment, we noted that about 7\% of  all Co atoms traced vanished after a voltage pulse, which we attribute to pick up by the tip. 
The probability for this pick up appears asymmetric with respect to the pulsing voltage.
For negative voltage pulses, the pick up probability was $22\pm 3$\%, against $9 \pm 2$\% for positive pulses.
Apparently, pulses with negative bias are more effective in picking up Co ad-atoms.
\\

The Co atoms sitting on the  Au substrate in the vicinity of the GNR occasionally also move to a new position after a voltage pulse. 
This motivated us to perform additional experiments to also study Co atom migration 
resulting from voltage pulses applied on the bare Au surface\cite{Braun2007,Fernandez2006}  in comparison to 
applying voltage pulses directly on the GNR.  
To this end we used the same initial tunnel current settings and a very similar procedure of lowering
the tip  as for pulses applied on GNRs, where we applied voltage pulses in the range of $-1000$\,mV$\le V_{\rm max} \le - 600$\,mV and 
$+600$\,mV$\le V_{\rm max} \le +1200$\,mV. In both cases we restrict the analysis to atoms within a distance of 7\,nm from the tip, 
because this covers most events, and it is an area that is visible in most of the scans taken. Within this range we find that 391/6331 (6\,\%)
of Co atoms on the Au surface are displaced as the result of a pulse on a neighboring GNR. For pulses directly on Au we find 16 events 
for 519 atoms in the range (3\,\%). Although it appears that applying voltage pulses on GNRs is slightly more effective we are not certain
about the statistical significance, or the role of the small differences in forming contacts on Au vs GNR. { However, a clear
difference is observed when comparing the effect of the pulses on Co ad-atoms on the GNR, where 694/1430 are displaced within
this same range (49\,\%), for pulses applied on the GNR. }

It will be useful to compare the effect of voltage-pulse induced atom migration to thermal activation. For this purpose we performed a series of experiments at 
elevated temperatures, in analogy to experiments reported previously.\cite{Repp2016} This was done by heating of the STM head, which is thermally weakly coupled to the helium bath. 
The temperature was recorded by a DT-470 diode sensor (Lake Shore Cryotronics), and we waited several hours for the temperature to stabilize before
starting the measurements. At each temperature, sequences of STM images of the same surface area were recorded. By comparing subsequent images 
we analyze the diffusion of the ad-atoms, where the positions of the GNRs and the herringbone structure served as immobile anchors for drift correction.  

The temperature was raised from the starting value of 8\,K, in steps of $\sim15$\,K. 
Thermally activated displacement of Co atoms on the bare gold surface was first detected when the temperature reached 34\,K. 
At still higher temperatures the speed of diffusion complicates meaningful quantitative analysis of the images.  
However, even at our highest temperature, 79\,K, motion of Co on GNRs was not observed. 
This experiment demonstrates that the diffusion barrier for Co ad-atoms on GNRs is much 
higher than that for Co on Au(111). 
Based on the observed diffusion rate of Co/Au(111) at 34\,K, and assuming an attempt frequency that is typically in order of $\nu_0 = 10^{13}$\,s$^{-1}$, 
we estimate the diffusion barrier of Co on Au(111) to be $E_D^\textrm{Au} = 0.10$\,eV in agreement with a theoretical study\cite{Bulou2008}.
{ As mentioned above, for current injection we observed a rather low escape rate for Co atoms directly adsorbed on Au, indicating that current-induced heating keeps the temperature of the gold surface
well below 40\,K. This is in agreement with our estimates based on the large thermal conductivity of bulk Au.}

We now turn to a discussion of the observed ad-atom hopping on GNRs. 
The first remarkable observation is the fact that, upon activation by a voltage pulse, the Co atoms diffuse along the GNR and only drop to the Au surface in exceptional cases.
This suggests the presence of an Ehrlich-Schwoebel barrier at the edge. 
In addition, the potential at the ends of the GNR, and at the kinks, must show a deep
minimum, because we find that Co atoms remain stationary at these points. 
Apart from these special points we observe diffusion of Co atoms along the GNR, which is nearly 1D.
Since we do not detect any preferential direction of motion it is reasonable to assume 
that the motion is activated by current-induced heating of the GNR.\cite{Schulze2008a,Schulze2008b,Neel2008,Ye2016,Erpenbeck2018,Preston2020}  
Adopting a theoretical value of about $E_D^{\rm GNR} \approx 0.5$\,eV for the diffusion barrier for 
Co on single layer graphene from theory (see Ref.~\citenum{Manade2015}, and references therein), 
we can use the observed hopping rate  $\nu$ during a pulse,  $\nu \sim  25$\,s$^{-1}$,  
for estimating the effective temperature on the GNR. Here, we have calculated $\nu$ from the average number of hopping events,
$N=13$, obtained above from the fit to Fig.~\ref{Fig:histogram}, and the typical pulse duration of 0.5\,s. Through the Arrhenius law for hopping 
$\nu = \nu_0\exp\left(-E_D / (k_BT)\right)$
we obtain an effective temperature $T_{\rm eff}\simeq200$\,K, which is indeed much higher 
than the temperatures that we were able to explore by heating of the STM. 
{Indeed, arguing from the opposite side}, at the highest temperature of 79\,K set by heating, a hopping 
barrier of 0.5\,eV inhibits any hopping at reasonable timescales. We estimate the error in the effective temperature as 20\,K.

Provided that the current-induced diffusion can be regarded as analogous to thermally driven diffusion,\cite{Meair2014} 
our observations can be interpreted as follows.
The fact that the escape probability, Fig.~\ref{Fig:escape_probability}(a), is nearly constant up to $\sim 8$\,nm, and only drops by a factor of $\sim 4$ at 20\,nm, 
suggests that the effective temperature of the GNR during a pulse is raised above that 
of the Au substrate, and is nearly homogeneous over long distances along its length. 
The exponential dependence of hopping rate with temperature permits a drop of the
effective temperature by only about 10\,K at the largest distance $s = 20$\,nm, with respect to the estimated 
pulse-induced temperature of $T_{\rm eff}=200$\,K. 
Let us verify whether this is consistent with known thermal transport properties. If we adopt the simplifying assumption that the Au substrate remains uniformly  
at the base temperature, the balance between thermal transport along the GNR, and heat transport into the substrate, for an infinitely long GNR leads to a decay of temperature 
with distance $s$ as, $T(s) = T_m \exp(-s/\lambda) + T_0$. Here, $T_m$ is the maximum temperature at the position of the tip, and the 
decay length is given by $\lambda = \sqrt{\kappa b/G}$. For the thermal conductivity of graphene 
we adopt the value $\kappa =200$\,W/Km,\cite{Xu2014,Guo2009} appropriate for 200\, K and for the typical length of 10--20\,nm of our GNRs. 
For the thickness $b = 0.144$\,nm of graphene this is equivalent to a two dimensional thermal conductivity of 
$\kappa_{\Box} = 30$\,nW/K.
For the Kapitza interface conductivity we take the value for the interface between graphite and Au,\cite{Schmidt2010} 
$G = 30\cdot 10^6$\,W/m$^2$K. 
From these two competing thermal conductivities 
we find a decay length of $\lambda = 31$\,nm for the temperature along the ribbons. 
Note that this is an under-estimate
because we have assumed that the GNR is infinitely long. Therefore, this estimate {supports the assessment} that the temperature 
of the GNR should be nearly constant along the GNR during a pulse.

Turning to the distribution of hopping distances in Fig.~\ref{Fig:histogram}, in case of thermal activation we expect the distribution to be closely approximated by that for a 1D random walk,
for which the probability for an atom to move by $n$ lattice sites, $P(n)$,  in the limit of large numbers of jumps, $N$, takes the form, 
\begin{equation}
P(n) = \sqrt{\frac{2}{N\pi}}\, e^{-\frac{n^2}{2N}}.
\label{eq.random-walk}
\end{equation}
The histogram in Fig.~\ref{Fig:histogram} agrees with such a description for short hopping distances, but the large peak at $d=0$  and the large displacements up to $d = 8$\,nm are clearly incompatible
with this description. Likely, these deviations result from the variations in effective temperatures of the GNR between pulses. The currents resulting from a voltage pulse vary because of variations in the 
contact resistance between tip and GNR. In addition, the total length of the GNR is expected to influence the effective temperature, because longer GNRs have a larger interface with the Au substrate for thermal relaxation. The diffusion rate is exponentially sensitive to variations in the effective temperature. E.g., when the effective temperature drops by only 20\,K the average number of 
hops during a pulse drops below 1, which we present as an interpretation of the large peak at $d=0$. Similarly, an effective temperature that is 20\,K higher than the estimate of $T_{\rm eff}\simeq 200$\,K given above is compatible with hopping to distances as large as 8\,nm.  Despite these uncertainties, the global behavior is in agreement with thermally activated 1D diffusion. 

The view that missing ad-atoms have been picked up by the tip is supported by the observation that the probability for atoms to be removed is asymmetric in the polarity of the tip voltage. 
The pick up of Co atoms is promoted by the electric field present in the junction, in combination with the van der Waals force of the tip acting on the ad-atoms.
For both voltage polarities, the van der Waals force is the same. 
The electric field, on the other hand, is determined by the applied bias voltage, by the difference in work functions of tip and sample, and by the effective charge on the ad-atom.
In order to determine the difference in work function we employed Kelvin probe force spectroscopy, 
from which we conclude that the work function of the Au(111) sample is 0.56\,eV higher than the one for the tip. 
In addition, charge transfer between the GNR and the Co atom results in an effective positive charge on Co.\cite{Manade2015}
These two effects add up to enhance the force pulling the atom towards the tip for negative voltage pulses, and reduce the force at the opposite polarity. 

While we observe a tip polarity dependence in the probability for ad-atom pick-up, we find no polarity dependence in ad-atom diffusion along the GNR. 
Directional asymmetry could result from electromigration forces. Here, we conventionally distinguish a direct force and a wind force.\cite{Sorbello1998}
The direct force arises from the effective charge of the Co ad-atom in the electric field at the surface. The horizontal component of the field
should be very small, because of the large effective tip radius, of order 50\,nm, and the boundary conditions to the field imposed by
the Au surface.\footnote{Note that the effective tip radius relevant for tunneling is different from the tip radius for electrostatics.}
 For an ad-atom sitting 0.3\,nm above the image plane, at a distance $s=10$\,nm from the point of tunneling, and for a tip-sample potential difference of 1V, 
the horizontal component of the electric field can be estimated as $2\cdot 10^6$\,V/m. For a charge of 0.5\,e on the ad-atom\cite{Manade2015}, 
the force would be 0.16\,pN. The resulting difference in left and right barrier heights would be $\Delta E = 0.2$\,meV, and at the induced temperature
of $\sim 200$\,K the ratio of left and right jumps would be 0.988. With the average number of jumps of $N=13$ this would lead to an expected average displacement of $\bar{d}=0.16$\,nm.
From the experiments we obtain $2\bar{d}=0.1\pm 0.2$\,nm as the difference between the mean displacement for the two voltage polarities. Although this shows that we cannot resolve the effect from
our data, the analysis suggests that the directional drift component should be detectable by improving the statistics. 

The wind force is more difficult to estimate, but is expected to be very small in our experiment. The 7-aGNR has a band gap of about 2.5\,eV, and the low-energy density of states  for 7-aGNR
is dominated by Au-derived states.\cite{Deniz2017} Although a part of the current may be carried along the GNR by the valence band states for $V < -0.8$\,V, the current will predominantly pass through the GNR directly to the Au substrate.  

In summary, we have shown that 7-aGNR on Au(111) serve as nearly perfect 1D channels for Co ad-atom diffusion. 
Current pulses injected from an STM tip can be used to heat the GNR, and the statistics of ad-atom diffusion suggests that this leads to a nearly homogeneous temperature of the GNR. 
This temperature is remarkably high, and far from equilibrium with the underlying Au substrate.
Further elaboration of this platform of STM observation of ad-atom diffusion on GNRs can serve as a nearly ideal model system for the study of electromigration at the atomic scale. 
Parameters obtained from computational studies for ad-atoms on graphene suggest that the experiment can be set up to achieve the required current densities.\cite{Solenov2012} 
Relevant adjustements would include replacing our 7-aGNR by nanoribbons with a small band gap, such as 5-aGNR,\cite{Kimouche2015,Lawrence2020} or 9-aGNR,\cite{Talirz2017} 
and arranging the GNR over the edge of thin insulating islands.\cite{Jacobse2018} 

\begin{acknowledgement}
This work is part of the research program of the Netherland's Organization for Scientific Research, NWO. Funding from the Deutsche Forschungsgemeinschaft (DFG, German Research Foundation) through SFB 689, project A07 and 314695032 / CRC 1277, projects A09 and B01 is gratefully acknowledged.
\end{acknowledgement}

\bibliography{lit_em}

\providecommand{\latin}[1]{#1}
\makeatletter
\providecommand{\doi}
  {\begingroup\let\do\@makeother\dospecials
  \catcode`\{=1 \catcode`\}=2\doi@aux}
\providecommand{\doi@aux}[1]{\endgroup\texttt{#1}}
\makeatother
\providecommand*\mcitethebibliography{\thebibliography}
\csname @ifundefined\endcsname{endmcitethebibliography}
  {\let\endmcitethebibliography\endthebibliography}{}
\begin{mcitethebibliography}{46}
\providecommand*\natexlab[1]{#1}
\providecommand*\mciteSetBstSublistMode[1]{}
\providecommand*\mciteSetBstMaxWidthForm[2]{}
\providecommand*\mciteBstWouldAddEndPuncttrue
  {\def\EndOfBibitem{\unskip.}}
\providecommand*\mciteBstWouldAddEndPunctfalse
  {\let\EndOfBibitem\relax}
\providecommand*\mciteSetBstMidEndSepPunct[3]{}
\providecommand*\mciteSetBstSublistLabelBeginEnd[3]{}
\providecommand*\EndOfBibitem{}
\mciteSetBstSublistMode{f}
\mciteSetBstMaxWidthForm{subitem}{(\alph{mcitesubitemcount})}
\mciteSetBstSublistLabelBeginEnd
  {\mcitemaxwidthsubitemform\space}
  {\relax}
  {\relax}

\bibitem[Cai \latin{et~al.}(2010)Cai, Ruffieux, Jaafar, Bieri, Braun,
  Blankenburg, Muoth, Seitsonen, Saleh, Feng, M\"ullen, and Fasel]{Cai2010}
Cai,~J.; Ruffieux,~P.; Jaafar,~R.; Bieri,~M.; Braun,~T.; Blankenburg,~S.;
  Muoth,~M.; Seitsonen,~A.~P.; Saleh,~M.; Feng,~X.; M\"ullen,~K.; Fasel,~R.
  Atomically precise bottom-up fabrication of graphene nanoribbons.
  \emph{Nature} \textbf{2010}, \emph{466}, 470--473\relax
\mciteBstWouldAddEndPuncttrue
\mciteSetBstMidEndSepPunct{\mcitedefaultmidpunct}
{\mcitedefaultendpunct}{\mcitedefaultseppunct}\relax
\EndOfBibitem
\bibitem[Rizzo \latin{et~al.}(2018)Rizzo, Veber, Cao, Bronner, Chen, Zhao,
  Rodriguez, Louie, Crommie, and Fischer]{Rizzo2018}
Rizzo,~D.~J.; Veber,~G.; Cao,~T.; Bronner,~C.; Chen,~T.; Zhao,~F.;
  Rodriguez,~H.; Louie,~S.~G.; Crommie,~M.~F.; Fischer,~F.~R. Topological band
  engineering of graphene nanoribbons. \emph{Nature} \textbf{2018}, \emph{560},
  204--208\relax
\mciteBstWouldAddEndPuncttrue
\mciteSetBstMidEndSepPunct{\mcitedefaultmidpunct}
{\mcitedefaultendpunct}{\mcitedefaultseppunct}\relax
\EndOfBibitem
\bibitem[Gr\"oning \latin{et~al.}(2018)Gr\"oning, Wang, Yao, Pignedoli,
  Borin~Barin, Daniels, Cupo, Meunier, Feng, Narita, M\"ullen, Ruffieux, and
  Fasel]{Groning2018}
Gr\"oning,~O.; Wang,~S.; Yao,~X.; Pignedoli,~C.~A.; Borin~Barin,~G.;
  Daniels,~C.; Cupo,~A.; Meunier,~V.; Feng,~X.; Narita,~A.; M\"ullen,~K.;
  Ruffieux,~P.; Fasel,~R. Engineering of robust topological quantum phases in
  graphene nanoribbons. \emph{Nature} \textbf{2018}, \emph{560}, 209--213\relax
\mciteBstWouldAddEndPuncttrue
\mciteSetBstMidEndSepPunct{\mcitedefaultmidpunct}
{\mcitedefaultendpunct}{\mcitedefaultseppunct}\relax
\EndOfBibitem
\bibitem[Slota \latin{et~al.}(2018)Slota, Keerthi, Myers, Tretyakov,
  Baumgarten, Ardavan, Sadeghi, Lambert, Narita, M\"ullen, and
  Bogani]{Slota2018}
Slota,~M.; Keerthi,~A.; Myers,~W.~K.; Tretyakov,~E.; Baumgarten,~M.;
  Ardavan,~A.; Sadeghi,~H.; Lambert,~C.~J.; Narita,~A.; M\"ullen,~K.;
  Bogani,~L. Magnetic edge states and coherent manipulation of graphene
  nanoribbons. \emph{Nature} \textbf{2018}, \emph{557}, 691--695\relax
\mciteBstWouldAddEndPuncttrue
\mciteSetBstMidEndSepPunct{\mcitedefaultmidpunct}
{\mcitedefaultendpunct}{\mcitedefaultseppunct}\relax
\EndOfBibitem
\bibitem[Nguyen \latin{et~al.}(2017)Nguyen, Tsai, Omrani, Marangoni, Wu, Rizzo,
  Rodgers, Cloke, Durr, Sakai, Liou, Aikawa, Chelikowsky, Louie, Fischer, and
  Crommie]{Nguyen2017}
Nguyen,~G.~D. \latin{et~al.}  Atomically precise graphene nanoribbon
  heterojunctions from a single molecular precursor. \emph{Nat. Nanotechnol.}
  \textbf{2017}, \relax
\mciteBstWouldAddEndPunctfalse
\mciteSetBstMidEndSepPunct{\mcitedefaultmidpunct}
{}{\mcitedefaultseppunct}\relax
\EndOfBibitem
\bibitem[Jacobse \latin{et~al.}(2017)Jacobse, Kimouche, Gebraad, Ervasti,
  Thijssen, Liljeroth, and Swart]{Jacobse2017}
Jacobse,~P.~H.; Kimouche,~A.; Gebraad,~T.; Ervasti,~M.~M.; Thijssen,~J.~M.;
  Liljeroth,~P.; Swart,~I. Electronic components embedded in a single graphene
  nanoribbon. \emph{Nat. Commun.} \textbf{2017}, \emph{8}, 119\relax
\mciteBstWouldAddEndPuncttrue
\mciteSetBstMidEndSepPunct{\mcitedefaultmidpunct}
{\mcitedefaultendpunct}{\mcitedefaultseppunct}\relax
\EndOfBibitem
\bibitem[Koch \latin{et~al.}(2012)Koch, Ample, Joachim, and Grill]{Koch2012}
Koch,~M.; Ample,~F.; Joachim,~C.; Grill,~L. Voltage-dependent conductance of a
  single graphene nanoribbon. \emph{Nat. Nanotechnol.} \textbf{2012}, \emph{7},
  713--717\relax
\mciteBstWouldAddEndPuncttrue
\mciteSetBstMidEndSepPunct{\mcitedefaultmidpunct}
{\mcitedefaultendpunct}{\mcitedefaultseppunct}\relax
\EndOfBibitem
\bibitem[Jacobse \latin{et~al.}(2018)Jacobse, Mangnus, Zevenhuizen, and
  Swart]{Jacobse2018}
Jacobse,~P.~H.; Mangnus,~M. J.~J.; Zevenhuizen,~S. J.~M.; Swart,~I. Mapping the
  Conductance of Electronically Decoupled Graphene Nanoribbons. \emph{ACS Nano}
  \textbf{2018}, \emph{12}, 7048--7056\relax
\mciteBstWouldAddEndPuncttrue
\mciteSetBstMidEndSepPunct{\mcitedefaultmidpunct}
{\mcitedefaultendpunct}{\mcitedefaultseppunct}\relax
\EndOfBibitem
\bibitem[Li \latin{et~al.}(2013)Li, Zhang, Morgenstern, and Mazzarello]{Li2013}
Li,~Y.; Zhang,~W.; Morgenstern,~M.; Mazzarello,~R. Electronic and Magnetic
  Properties of Zigzag Graphene Nanoribbons on the (111) Surface of Cu, Ag, and
  Au. \emph{Phys. Rev. Lett.} \textbf{2013}, \emph{110}, 216804\relax
\mciteBstWouldAddEndPuncttrue
\mciteSetBstMidEndSepPunct{\mcitedefaultmidpunct}
{\mcitedefaultendpunct}{\mcitedefaultseppunct}\relax
\EndOfBibitem
\bibitem[Ij\"as \latin{et~al.}(2013)Ij\"as, Ervasti, Uppstu, Liljeroth, van~der
  Lit, Swart, and Harju]{Ijaes2013}
Ij\"as,~M.; Ervasti,~M.; Uppstu,~A.; Liljeroth,~P.; van~der Lit,~J.; Swart,~I.;
  Harju,~A. Electronic states in finite graphene nanoribbons: Effect of
  charging and defects. \emph{Phys. Rev. B} \textbf{2013}, \emph{88},
  075429\relax
\mciteBstWouldAddEndPuncttrue
\mciteSetBstMidEndSepPunct{\mcitedefaultmidpunct}
{\mcitedefaultendpunct}{\mcitedefaultseppunct}\relax
\EndOfBibitem
\bibitem[van~der Lit \latin{et~al.}(2013)van~der Lit, Boneschanscher,
  Vanmaekelbergh, Ij\"as, Uppstu, Ervasti, Harju, Liljeroth, and
  Swart]{vanderLit2013}
van~der Lit,~J.; Boneschanscher,~M.~P.; Vanmaekelbergh,~D.; Ij\"as,~M.;
  Uppstu,~A.; Ervasti,~M.; Harju,~A.; Liljeroth,~P.; Swart,~I. Suppression of
  electron-vibron coupling in graphene nanoribbons contacted via a single atom.
  \emph{Nat. Commun.} \textbf{2013}, \emph{4}\relax
\mciteBstWouldAddEndPuncttrue
\mciteSetBstMidEndSepPunct{\mcitedefaultmidpunct}
{\mcitedefaultendpunct}{\mcitedefaultseppunct}\relax
\EndOfBibitem
\bibitem[Baringhaus \latin{et~al.}(2013)Baringhaus, Edler, Neumann, Stampfer,
  Forti, Starke, and Tegenkamp]{Baringhaus2013}
Baringhaus,~J.; Edler,~F.; Neumann,~C.; Stampfer,~C.; Forti,~S.; Starke,~U.;
  Tegenkamp,~C. Local transport measurements on epitaxial graphene. \emph{Appl.
  Phys. Lett.} \textbf{2013}, \emph{103}, 111604\relax
\mciteBstWouldAddEndPuncttrue
\mciteSetBstMidEndSepPunct{\mcitedefaultmidpunct}
{\mcitedefaultendpunct}{\mcitedefaultseppunct}\relax
\EndOfBibitem
\bibitem[Kawai \latin{et~al.}(2015)Kawai, Saito, Osumi, Yamaguchi, Foster,
  Spijker, and Meyer]{Kawai2015}
Kawai,~S.; Saito,~S.; Osumi,~S.; Yamaguchi,~S.; Foster,~A.~S.; Spijker,~P.;
  Meyer,~E. Atomically controlled substitutional boron-doping of graphene
  nanoribbons. \emph{Nat. Commun.} \textbf{2015}, \emph{6}, 8098\relax
\mciteBstWouldAddEndPuncttrue
\mciteSetBstMidEndSepPunct{\mcitedefaultmidpunct}
{\mcitedefaultendpunct}{\mcitedefaultseppunct}\relax
\EndOfBibitem
\bibitem[Carbonell-Sanrom\`{a} \latin{et~al.}(2017)Carbonell-Sanrom\`{a},
  Hieulle, Vilas-Varela, Brandimarte, Iraola, Barrag\'an, Li, Abadia, Corso,
  S\'anchez-Portal, Pe{\~{n}}a, and Pascual]{Carbonell-Sanroma2017b}
Carbonell-Sanrom\`{a},~E.; Hieulle,~J.; Vilas-Varela,~M.; Brandimarte,~P.;
  Iraola,~M.; Barrag\'an,~A.; Li,~J.; Abadia,~M.; Corso,~M.;
  S\'anchez-Portal,~D.; Pe{\~{n}}a,~D.; Pascual,~J.~I. Doping of Graphene
  Nanoribbons via Functional Group Edge Modification. \emph{ACS Nano}
  \textbf{2017}, \emph{11}, 7355--7361\relax
\mciteBstWouldAddEndPuncttrue
\mciteSetBstMidEndSepPunct{\mcitedefaultmidpunct}
{\mcitedefaultendpunct}{\mcitedefaultseppunct}\relax
\EndOfBibitem
\bibitem[Chen \latin{et~al.}(2008)Chen, Jang, Adam, Fuhrer, Williams, and
  Ishigami]{Chen2008}
Chen,~J.-H.; Jang,~C.; Adam,~S.; Fuhrer,~M.~S.; Williams,~E.~D.; Ishigami,~M.
  Charged-impurity scattering in graphene. \emph{Nature Physics} \textbf{2008},
  \emph{4}, 377--381\relax
\mciteBstWouldAddEndPuncttrue
\mciteSetBstMidEndSepPunct{\mcitedefaultmidpunct}
{\mcitedefaultendpunct}{\mcitedefaultseppunct}\relax
\EndOfBibitem
\bibitem[Elias and Henriksen(2020)Elias, and Henriksen]{Elias2020}
Elias,~J.; Henriksen,~E. Unexpected Hole Doping of Graphene by Osmium Adatoms.
  \emph{Ann. Phys. (Berlin)} \textbf{2020}, \emph{532}, 1900294\relax
\mciteBstWouldAddEndPuncttrue
\mciteSetBstMidEndSepPunct{\mcitedefaultmidpunct}
{\mcitedefaultendpunct}{\mcitedefaultseppunct}\relax
\EndOfBibitem
\bibitem[Dundas \latin{et~al.}(2009)Dundas, McEniry, and Todorov]{Dundas2009}
Dundas,~D.; McEniry,~E.~J.; Todorov,~T.~N. Current-driven atomic waterwheels.
  \emph{Nat. Nanotechnol.} \textbf{2009}, \emph{4}, 99--102\relax
\mciteBstWouldAddEndPuncttrue
\mciteSetBstMidEndSepPunct{\mcitedefaultmidpunct}
{\mcitedefaultendpunct}{\mcitedefaultseppunct}\relax
\EndOfBibitem
\bibitem[L{\"u} \latin{et~al.}(2010)L{\"u}, Brandbyge, and
  Hedeg{\aa}rd]{Lu2010}
L{\"u},~J.-T.; Brandbyge,~M.; Hedeg{\aa}rd,~P. Blowing the Fuse: Berry's Phase
  and Runaway Vibrations in Molecular Conductors. \emph{Nano Lett.}
  \textbf{2010}, \emph{10}, 1657--1663\relax
\mciteBstWouldAddEndPuncttrue
\mciteSetBstMidEndSepPunct{\mcitedefaultmidpunct}
{\mcitedefaultendpunct}{\mcitedefaultseppunct}\relax
\EndOfBibitem
\bibitem[Dienel \latin{et~al.}(2015)Dienel, Kawai, S\"ode, Feng, M\"ullen,
  Ruffieux, Fasel, and Gr\"oning]{Dienel2015}
Dienel,~T.; Kawai,~S.; S\"ode,~H.; Feng,~X.; M\"ullen,~K.; Ruffieux,~P.;
  Fasel,~R.; Gr\"oning,~O. Resolving Atomic Connectivity in Graphene
  Nanostructure Junctions. \emph{Nano Lett.} \textbf{2015}, \emph{15},
  5185--5190\relax
\mciteBstWouldAddEndPuncttrue
\mciteSetBstMidEndSepPunct{\mcitedefaultmidpunct}
{\mcitedefaultendpunct}{\mcitedefaultseppunct}\relax
\EndOfBibitem
\bibitem[Repp \latin{et~al.}(2004)Repp, Meyer, Olsson, and Persson]{Repp2004b}
Repp,~J.; Meyer,~G.; Olsson,~F.~E.; Persson,~M. Controlling the Charge State of
  Individual Gold Adatoms. \emph{Science} \textbf{2004}, \emph{305},
  493--495\relax
\mciteBstWouldAddEndPuncttrue
\mciteSetBstMidEndSepPunct{\mcitedefaultmidpunct}
{\mcitedefaultendpunct}{\mcitedefaultseppunct}\relax
\EndOfBibitem
\bibitem[Virgus \latin{et~al.}(2014)Virgus, Purwanto, Krakauer, and
  Zhang]{Virgus2014}
Virgus,~Y.; Purwanto,~W.; Krakauer,~H.; Zhang,~S. Stability, Energetics, and
  Magnetic States of Cobalt Adatoms on Graphene. \emph{Phys. Rev. Lett.}
  \textbf{2014}, \emph{113}, 175502\relax
\mciteBstWouldAddEndPuncttrue
\mciteSetBstMidEndSepPunct{\mcitedefaultmidpunct}
{\mcitedefaultendpunct}{\mcitedefaultseppunct}\relax
\EndOfBibitem
\bibitem[Sevin\c{c}li \latin{et~al.}(2008)Sevin\c{c}li, Topsakal, Durgun, and
  Ciraci]{Sevincli2008}
Sevin\c{c}li,~H.; Topsakal,~M.; Durgun,~E.; Ciraci,~S. Electronic and magnetic
  properties of $3d$ transition-metal atom adsorbed graphene and graphene
  nanoribbons. \emph{Phys. Rev. B} \textbf{2008}, \emph{77}, 195434\relax
\mciteBstWouldAddEndPuncttrue
\mciteSetBstMidEndSepPunct{\mcitedefaultmidpunct}
{\mcitedefaultendpunct}{\mcitedefaultseppunct}\relax
\EndOfBibitem
\bibitem[Untiedt \latin{et~al.}(2007)Untiedt, Caturla, Calvo, Palacios, Segers,
  and van Ruitenbeek]{Untiedt2007}
Untiedt,~C.; Caturla,~M.; Calvo,~R.; Palacios,~J.; Segers,~R.; van
  Ruitenbeek,~J. Formation of a metallic contact: jump to contact revisited.
  \emph{Phys. Rev. Lett.} \textbf{2007}, \emph{98}, 206801\relax
\mciteBstWouldAddEndPuncttrue
\mciteSetBstMidEndSepPunct{\mcitedefaultmidpunct}
{\mcitedefaultendpunct}{\mcitedefaultseppunct}\relax
\EndOfBibitem
\bibitem[Kr\"oger \latin{et~al.}(2007)Kr\"oger, Jensen, and
  Berndt]{Kroeger2007}
Kr\"oger,~J.; Jensen,~H.; Berndt, Conductance of tip-surface and tip-atom
  junctions on Au(111) explored by a scanning tunnelling microscope. \emph{New
  J. Phys.} \textbf{2007}, \emph{9}, 153\relax
\mciteBstWouldAddEndPuncttrue
\mciteSetBstMidEndSepPunct{\mcitedefaultmidpunct}
{\mcitedefaultendpunct}{\mcitedefaultseppunct}\relax
\EndOfBibitem
\bibitem[Braun \latin{et~al.}(2007)Braun, Soe, Flipse, and Rieder]{Braun2007}
Braun,~K.-F.; Soe,~W.-H.; Flipse,~C. F.~J.; Rieder,~K.-H. Electromigration of
  single metal atoms observed by scanning tunneling microscopy. \emph{Appl.
  Phys. Lett.} \textbf{2007}, \emph{90}, 023118\relax
\mciteBstWouldAddEndPuncttrue
\mciteSetBstMidEndSepPunct{\mcitedefaultmidpunct}
{\mcitedefaultendpunct}{\mcitedefaultseppunct}\relax
\EndOfBibitem
\bibitem[Fernandez-Torres \latin{et~al.}({2006})Fernandez-Torres, Sykes,
  Nanayakkara, and Weiss]{Fernandez2006}
Fernandez-Torres,~L.; Sykes,~E.; Nanayakkara,~S.; Weiss,~P. {Dynamics and
  spectroscopy of hydrogen atoms on Pd\{111\}}. \emph{J. Phys. Chem. B}
  \textbf{{2006}}, \emph{{110}}, {7380--7384}\relax
\mciteBstWouldAddEndPuncttrue
\mciteSetBstMidEndSepPunct{\mcitedefaultmidpunct}
{\mcitedefaultendpunct}{\mcitedefaultseppunct}\relax
\EndOfBibitem
\bibitem[Repp \latin{et~al.}(2016)Repp, Steurer, Scivetti, Persson, Gross, and
  Meyer]{Repp2016}
Repp,~J.; Steurer,~W.; Scivetti,~I.; Persson,~M.; Gross,~L.; Meyer,~G.
  Charge-State-Dependent Diffusion of Individual Gold Adatoms on Ionic Thin
  NaCl Films. \emph{Phys. Rev. Lett.} \textbf{2016}, \emph{117}, 146102\relax
\mciteBstWouldAddEndPuncttrue
\mciteSetBstMidEndSepPunct{\mcitedefaultmidpunct}
{\mcitedefaultendpunct}{\mcitedefaultseppunct}\relax
\EndOfBibitem
\bibitem[Bulou(2008)]{Bulou2008}
Bulou,~H. Atomic diffusion on nanostructured surfaces. \emph{Superlatt.
  Microstruct.} \textbf{2008}, \emph{44}, 533--541\relax
\mciteBstWouldAddEndPuncttrue
\mciteSetBstMidEndSepPunct{\mcitedefaultmidpunct}
{\mcitedefaultendpunct}{\mcitedefaultseppunct}\relax
\EndOfBibitem
\bibitem[Schulze \latin{et~al.}(2008)Schulze, Franke, and
  Pascual]{Schulze2008a}
Schulze,~G.; Franke,~K.~J.; Pascual,~J.~I. Resonant heating and
  substrate-mediated cooling of a single C60molecule in a tunnel junction.
  \emph{New J. Phys.} \textbf{2008}, \emph{10}, 065005\relax
\mciteBstWouldAddEndPuncttrue
\mciteSetBstMidEndSepPunct{\mcitedefaultmidpunct}
{\mcitedefaultendpunct}{\mcitedefaultseppunct}\relax
\EndOfBibitem
\bibitem[Schulze \latin{et~al.}(2008)Schulze, Franke, Gagliardi, Romano, Lin,
  Rosa, Niehaus, Frauenheim, Di~Carlo, Pecchia, and Pascual]{Schulze2008b}
Schulze,~G.; Franke,~K.~J.; Gagliardi,~A.; Romano,~G.; Lin,~C.~S.; Rosa,~A.~L.;
  Niehaus,~T.~A.; Frauenheim,~T.; Di~Carlo,~A.; Pecchia,~A.; Pascual,~J.~I.
  Resonant Electron Heating and Molecular Phonon Cooling in Single
  ${\mathrm{C}}_{60}$ Junctions. \emph{Phys. Rev. Lett.} \textbf{2008},
  \emph{100}, 136801\relax
\mciteBstWouldAddEndPuncttrue
\mciteSetBstMidEndSepPunct{\mcitedefaultmidpunct}
{\mcitedefaultendpunct}{\mcitedefaultseppunct}\relax
\EndOfBibitem
\bibitem[N\'eel \latin{et~al.}(2008)N\'eel, Limot, Kr\"oger, and
  Berndt]{Neel2008}
N\'eel,~N.; Limot,~L.; Kr\"oger,~J.; Berndt,~R. Rotation of {C}$_{60}$ in a
  single-molecule contact. \emph{Phys. Rev. B} \textbf{2008}, \emph{77},
  125431\relax
\mciteBstWouldAddEndPuncttrue
\mciteSetBstMidEndSepPunct{\mcitedefaultmidpunct}
{\mcitedefaultendpunct}{\mcitedefaultseppunct}\relax
\EndOfBibitem
\bibitem[Ye \latin{et~al.}(2016)Ye, Zheng, Yan, and Di~Ventra]{Ye2016}
Ye,~L.; Zheng,~X.; Yan,~Y.; Di~Ventra,~M. Thermodynamic meaning of local
  temperature of nonequilibrium open quantum systems. \emph{Phys. Rev. B}
  \textbf{2016}, \emph{94}, 245105\relax
\mciteBstWouldAddEndPuncttrue
\mciteSetBstMidEndSepPunct{\mcitedefaultmidpunct}
{\mcitedefaultendpunct}{\mcitedefaultseppunct}\relax
\EndOfBibitem
\bibitem[Erpenbeck \latin{et~al.}(2018)Erpenbeck, Schinabeck, Peskin, and
  Thoss]{Erpenbeck2018}
Erpenbeck,~A.; Schinabeck,~C.; Peskin,~U.; Thoss,~M. Current-induced bond
  rupture in single-molecule junctions. \emph{Phys. Rev. B} \textbf{2018},
  \emph{97}, 235452\relax
\mciteBstWouldAddEndPuncttrue
\mciteSetBstMidEndSepPunct{\mcitedefaultmidpunct}
{\mcitedefaultendpunct}{\mcitedefaultseppunct}\relax
\EndOfBibitem
\bibitem[Preston \latin{et~al.}(2020)Preston, Kershaw, and Kosov]{Preston2020}
Preston,~R.~J.; Kershaw,~V.~F.; Kosov,~D.~S. Current-induced atomic motion,
  structural instabilities, and negative temperatures on molecule-electrode
  interfaces in electronic junctions. \emph{Phys. Rev. B} \textbf{2020},
  \emph{101}, 155415\relax
\mciteBstWouldAddEndPuncttrue
\mciteSetBstMidEndSepPunct{\mcitedefaultmidpunct}
{\mcitedefaultendpunct}{\mcitedefaultseppunct}\relax
\EndOfBibitem
\bibitem[Manad\'e \latin{et~al.}(2015)Manad\'e, Vi{\~{n}}es, and
  Illas]{Manade2015}
Manad\'e,~M.; Vi{\~{n}}es,~F.; Illas,~F. Transition metal adatoms on graphene:
  A systematic density functional study. \emph{Carbon} \textbf{2015},
  \emph{95}, 525--534\relax
\mciteBstWouldAddEndPuncttrue
\mciteSetBstMidEndSepPunct{\mcitedefaultmidpunct}
{\mcitedefaultendpunct}{\mcitedefaultseppunct}\relax
\EndOfBibitem
\bibitem[Meair \latin{et~al.}(2014)Meair, Bergfield, Stafford, and
  Jacquod]{Meair2014}
Meair,~J.; Bergfield,~J.~P.; Stafford,~C.~A.; Jacquod,~P. Local temperature of
  out-of-equilibrium quantum electron systems. \emph{Phys. Rev. B}
  \textbf{2014}, \emph{90}, 035407\relax
\mciteBstWouldAddEndPuncttrue
\mciteSetBstMidEndSepPunct{\mcitedefaultmidpunct}
{\mcitedefaultendpunct}{\mcitedefaultseppunct}\relax
\EndOfBibitem
\bibitem[Xu \latin{et~al.}(2014)Xu, Pereira, Wang, Wu, Zhang, Zhao, Bae,
  Tinh~Bui, Xie, Thong, Hong, Loh, Donadio, Li, and {\"O}zyilmaz]{Xu2014}
Xu,~X.; Pereira,~L. F.~C.; Wang,~Y.; Wu,~J.; Zhang,~K.; Zhao,~X.; Bae,~S.;
  Tinh~Bui,~C.; Xie,~R.; Thong,~J. T.~L.; Hong,~B.~H.; Loh,~K.~P.; Donadio,~D.;
  Li,~B.; {\"O}zyilmaz,~B. Length-dependent thermal conductivity in suspended
  single-layer graphene. \emph{Nat. Commun.} \textbf{2014}, \emph{5},
  3689\relax
\mciteBstWouldAddEndPuncttrue
\mciteSetBstMidEndSepPunct{\mcitedefaultmidpunct}
{\mcitedefaultendpunct}{\mcitedefaultseppunct}\relax
\EndOfBibitem
\bibitem[Guo \latin{et~al.}(2009)Guo, Zang, and Gong]{Guo2009}
Guo,~Z.; Zang,~D.; Gong,~X.-G. Thermal conductivity of graphene nanoribbons.
  \emph{Appl. Phys. Lett.} \textbf{2009}, \emph{95}, 163103\relax
\mciteBstWouldAddEndPuncttrue
\mciteSetBstMidEndSepPunct{\mcitedefaultmidpunct}
{\mcitedefaultendpunct}{\mcitedefaultseppunct}\relax
\EndOfBibitem
\bibitem[Schmidt \latin{et~al.}(2010)Schmidt, Collins, Minnich, and
  Chen]{Schmidt2010}
Schmidt,~A.; Collins,~K.; Minnich,~A.; Chen,~G. Thermal conductance and phonon
  transmissivity of metal--graphite interfaces. \emph{J. Appl. Phys.}
  \textbf{2010}, \emph{107}, 104907\relax
\mciteBstWouldAddEndPuncttrue
\mciteSetBstMidEndSepPunct{\mcitedefaultmidpunct}
{\mcitedefaultendpunct}{\mcitedefaultseppunct}\relax
\EndOfBibitem
\bibitem[Sorbello \latin{et~al.}(1998)Sorbello, Ehrenreich, and
  Spaepen]{Sorbello1998}
Sorbello,~R.~S.; Ehrenreich,~H.; Spaepen,~F. Theory of Electromigration.
  \emph{Solid State Phys.} \textbf{1998}, \emph{51}, 159--231\relax
\mciteBstWouldAddEndPuncttrue
\mciteSetBstMidEndSepPunct{\mcitedefaultmidpunct}
{\mcitedefaultendpunct}{\mcitedefaultseppunct}\relax
\EndOfBibitem
\bibitem[Deniz \latin{et~al.}(2017)Deniz, S\'anchez-S\'anchez, Dumslaff, Feng,
  Narita, M\"ullen, Kharche, Meunier, Fasel, and Ruffieux]{Deniz2017}
Deniz,~O.; S\'anchez-S\'anchez,~C.; Dumslaff,~T.; Feng,~X.; Narita,~A.;
  M\"ullen,~K.; Kharche,~N.; Meunier,~V.; Fasel,~R.; Ruffieux,~P. Revealing the
  Electronic Structure of Silicon Intercalated Armchair Graphene Nanoribbons by
  Scanning Tunneling Spectroscopy. \emph{Nano Lett.} \textbf{2017}, \emph{17},
  2197--2203\relax
\mciteBstWouldAddEndPuncttrue
\mciteSetBstMidEndSepPunct{\mcitedefaultmidpunct}
{\mcitedefaultendpunct}{\mcitedefaultseppunct}\relax
\EndOfBibitem
\bibitem[Solenov and Velizhanin(2012)Solenov, and Velizhanin]{Solenov2012}
Solenov,~D.; Velizhanin,~K.~A. Adsorbate Transport on Graphene by
  Electromigration. \emph{Phys. Rev. Lett.} \textbf{2012}, \emph{109},
  095504\relax
\mciteBstWouldAddEndPuncttrue
\mciteSetBstMidEndSepPunct{\mcitedefaultmidpunct}
{\mcitedefaultendpunct}{\mcitedefaultseppunct}\relax
\EndOfBibitem
\bibitem[Kimouche \latin{et~al.}(2015)Kimouche, Ervasti, Drost, Halonen, Harju,
  Joensuu, Sainio, and Liljeroth]{Kimouche2015}
Kimouche,~A.; Ervasti,~M.~M.; Drost,~R.; Halonen,~S.; Harju,~A.;
  Joensuu,~P.~M.; Sainio,~J.; Liljeroth,~P. Ultra-narrow metallic armchair
  graphene nanoribbons. \emph{Nat. Commun.} \textbf{2015}, \emph{6},
  10177\relax
\mciteBstWouldAddEndPuncttrue
\mciteSetBstMidEndSepPunct{\mcitedefaultmidpunct}
{\mcitedefaultendpunct}{\mcitedefaultseppunct}\relax
\EndOfBibitem
\bibitem[Lawrence \latin{et~al.}(2020)Lawrence, Brandimarte, Berdonces-Layunta,
  Mohammed, Grewal, Leon, S\'anchez-Portal, and de~Oteyza]{Lawrence2020}
Lawrence,~J.; Brandimarte,~P.; Berdonces-Layunta,~A.; Mohammed,~M. S.~G.;
  Grewal,~A.; Leon,~C.~C.; S\'anchez-Portal,~D.; de~Oteyza,~D.~G. Probing the
  Magnetism of Topological End States in 5-Armchair Graphene Nanoribbons.
  \emph{ACS Nano} \textbf{2020}, \emph{14}, 4499--4508\relax
\mciteBstWouldAddEndPuncttrue
\mciteSetBstMidEndSepPunct{\mcitedefaultmidpunct}
{\mcitedefaultendpunct}{\mcitedefaultseppunct}\relax
\EndOfBibitem
\bibitem[Talirz \latin{et~al.}(2017)Talirz, S\"ode, Dumslaff, Wang,
  Sanchez-Valencia, Liu, Shinde, Pignedoli, Liang, Meunier, Plumb, Shi, Feng,
  Narita, M\"ullen, Fasel, and Ruffieux]{Talirz2017}
Talirz,~L. \latin{et~al.}  On-Surface Synthesis and Characterization of 9-Atom
  Wide Armchair Graphene Nanoribbons. \emph{ACS Nano} \textbf{2017}, \emph{11},
  1380--1388\relax
\mciteBstWouldAddEndPuncttrue
\mciteSetBstMidEndSepPunct{\mcitedefaultmidpunct}
{\mcitedefaultendpunct}{\mcitedefaultseppunct}\relax
\EndOfBibitem
\end{mcitethebibliography}

\end{document}